\documentstyle[aps,graphicx,epsf]{revtex}
\begin{document}

\title{Kaluza's theory in generalized coordinates.}

\author{Ana Laura Garc\'{i}a-Perciante$^*$, Alfredo Sandoval-Villalbazo $^{*\dag}$,
 and L.S. Garc\'{i}a Col\'{i}n.$^+$ }

\address{~}

\address{$^*$Departamento de Ciencias, Universidad
Iberoamericana, Col. Lomas de Santa Fe~01210, Mexico~D.F., Mexico}

\address{$^\dag$Relativity and Cosmology Group, School of
Computer Science and Mathematics, Portsmouth University,
Portsmouth~PO1~2EG, Britain. E-Mail: alfredo.sandoval@port.ac.uk}

\address{$^+$ Departamento de F\'{\i }sica,
Universidad Aut\'onoma Metropolitana-Iztapalapa,
Av. Pur\'{\i }sima y Michoac\'{a}n S/N. 
Mexico~D.F., 09340 Mexico. e-mail: lgcs@xanum.uam.mx}

\maketitle

\begin{abstract}
Maxwell's equations can be obtained in generalized coordinates by
considering the electromagnetic field as an external agent. The work here
presented shows how to obtain the electrodynamics for a charged particle in
generalized coordinates eliminating the concept of external force. Based on
Kaluza's formalism, the one here presented extends the 5x5 metric into a 6x6
space-time giving enough room to include magnetic monopoles in a very
natural way.
\end{abstract}

\smallskip

\section{Introduction}

Over seventy years ago, T. Kaluza developed a theory unifying
electromagnetism and gravitation by working in a five dimensional manifold 
\cite{Kaluza} \cite{Straumann}. In his work, influences of an electrical
charge in space are treated as sources of curvature in a similar way as mass
does in Einstein's theory of general relativity. With this formalism,
Maxwell's equations can be obtained in cartesian coordinates in a rather
simple way. In generalized coordinates, only four out of eight (three for
each rotational and one for each divergence) Maxwell equations can be
recovered correctly, those with sources. The problem resides in the
asymmetrical way in which each pair of equations, with and without sources,
are obtained {\em if Kaluza's formalism is applied}. The equations with
sources are obtained by means of the field equation, while the homogeneous
ones have to be deduced from an identity which yields incorrect results
unless the metric is cartesian. A more conventional method for obtaining
Maxwell equations consists in using Bianchi's identities\cite{Mesiner}.
Although it yields correct results using any metric, the electromagnetic
field is treated as an external ''source'' by including it in the stress
tensor. Kaluza's method, and consequently the one presented here, have the
advantage of considering charge as a space curvature thus following the
tenets of general relativity.

In this work we propose an alternative method for treating charge and mass
as curvature sources. This method is not only consistent with Kaluza's but
also represents a generalization of it, by considering his 5x5 metric
included in a larger one. A larger space also has enough room to establish a
generalized theory in which magnetic monopoles can be introduced in a very
natural way.

This paper is divided as follows. Section 2 contains a brief summary of
Kaluza's theory without specifying any metric, as presented in his original
article. In section 3, through a simplified example, the cause for the
incorrectness of the inhomogeneous equations is explored. Section 4 proposes
an alternative to the theory making a symmetrical generalization of the
metric used before, as a way of obtaining Maxwell equations using only
Einstein's field equation. Section 5 goes deeper into the interpretation of
the 6x6 space-time and finally in section 6 we discuss the implications of
this formalism.

\section{\bf Kaluza's theory}

Kaluza's formalism generates Maxwell's equations as well as the equations of
motion for charged particles by working with a 5x5 metric defined as\vspace{%
1pt}

\begin{equation}
g_{\mu \nu }=\left[ 
\begin{array}{lllll}
g_{11} & g_{12} & g_{13} & g_{14} & A_{1} \\ 
g_{21} & g_{22} & g_{23} & g_{24} & A_{2} \\ 
g_{31} & g_{32} & g_{33} & g_{34} & A_{3} \\ 
g_{41} & g_{42} & g_{43} & g_{44} & \frac{1}{c}\phi \\ 
A_{1} & A_{2} & A_{3} & \frac{1}{c}\phi & g_{55}
\end{array}
\right]  \label{met5gen}
\end{equation}
The 4x4 metric consisting of $g_{\mu \nu }$ elements with subscripts running
from 1 to 4 has to be a solution of Einstein's equations, while the fifth
column and row contain the four-vector electromagnetic potential $A_{\nu },$
the first three components being the ones of the usual potential vector and
the fourth the scalar potential $\phi $ . The fifth element $g_{55}$ is
undefined in Kaluza's article, but, as this extra dimension is considered as
a spatial type one, $g_{55}$ can be set equal to a constant taken to be
equal to a one in the case of a Minkowski metric.\vspace{1pt} \ \ \ \ \ \ \
\ \ \ \ \ \ \ \ \ \ \ \ \ \ \ \ \ \ \ \ \ \ \ \ \ \ \ \ \ \ \ \ \ \ \ \ \ \
\ \ \ \ \ \ \ \ \ \ \ \ \ \ \ \ \ \ \ \ \ \ \ \ \ \ \ \ \ \ \ \ \ \ \ \ \ \
\ \ \ \ \ \ \ \ \ \ \ \ \ \ \ 

\ The position and velocity vectors to consider in this formalism are
defined as follows

\begin{equation}
x^{\nu }=\left[ 
\begin{array}{l}
x^{1} \\ 
x^{2} \\ 
x^{3} \\ 
ct \\ 
x^{5}
\end{array}
\right]  \label{pos5gen}
\end{equation}

\begin{equation}
v^{\nu }=\left[ 
\begin{array}{l}
v^{1} \\ 
v^{2} \\ 
v^{3} \\ 
c \\ 
\frac{q}{m}
\end{array}
\right]  \label{vel5gen}
\end{equation}
The fact that the time derivative of the fifth position coordinate is taken
to be equal to the charge-mass ratio arises from comparing the equation of
motion for a charged particle moving under Lorenz's force

\begin{equation}
\frac{d^{2}x^{\alpha }}{dt^{2}}=\frac{q}{m}\left[ \varepsilon _{\beta \gamma
}^{\alpha }\frac{\partial x^{\alpha }}{\partial t}B^{\gamma }+E^{\gamma
}\right]  \label{lor5gen}
\end{equation}
with the one obtained by restricting the particle to move trough a geodesic
in this space namely,

\begin{equation}
\frac{d^{2}x^{\alpha }}{ds^{2}}+\Gamma _{\mu \nu }^{\alpha }\frac{dx^{\mu }}{%
ds}\frac{dx^{\nu }}{ds}=0  \label{geod}
\end{equation}
Indeed Eq. (\ref{lor5gen}) may be obtained from equation \ref{geod} only
if $v^{5}=\frac{q}{m}$. An extra condition must be taken into account in
order to recover electrodynamics within this framework. Such condition
usually referred to as the cylindrical condition \cite{Kaluza} namely,

\begin{equation}
\frac{\partial }{\partial x^{5}}=0  \label{cil5}
\end{equation}
makes any derivative with respect to the fifth component equal to cero. The
only justification for this restriction is that it leads to the correct
Maxwell equations.

The structural similarity between curls and Christoffel symbols, $\Gamma
_{\mu \nu }^{\alpha }$, suggest a proportionality between the latter and the
components of the electromagnetic field\cite{Kaluza}. These symbols, taking $%
g^{ij}=\delta ^{ij}$ (latin indices run from 1 to 3) are computed as follows

\begin{equation}
\Gamma _{\mu \nu }^{\alpha }=\frac{1}{2}g^{\alpha \lambda }\left( \frac{%
\partial g_{\mu \lambda }}{\partial x^{\nu }}+\frac{\partial g_{\nu \lambda }%
}{\partial x^{\mu }}-\frac{\partial g_{\mu \nu }}{\partial x^{\lambda }}%
\right)  \label{chrisgen}
\end{equation}

\begin{equation}
\Gamma _{\mu 5}^{\alpha }=\Gamma _{5\mu }^{\alpha }=\frac{1}{2}\left( \frac{%
\partial g_{5\lambda }}{\partial x^{\mu }}-\frac{\partial g_{\mu 5}}{%
\partial x^{\lambda }}\right) =F_{\mu }^{\alpha }  \label{chris5gen}
\end{equation}
where $F_{\mu }^{\alpha }$ , the elements of the field tensor, are defined
as follows

\begin{equation}
F_{\mu }^{\alpha }=\left[ 
\begin{array}{llll}
0 & B_{z} & -B_{y} & -\frac{1}{c}E_{x} \\ 
-B_{z} & 0 & B_{x} & -\frac{1}{c}E_{y} \\ 
B_{y} & -B_{x} & 0 & -\frac{1}{c}E_{z} \\ 
\frac{1}{c}E_{x} & \frac{1}{c}E_{y} & \frac{1}{c}E_{z} & 0
\end{array}
\right]  \label{faraday}
\end{equation}
In this scheme, Maxwell inhomogeneous equations may be obtained trough
Einstein's field equation

\begin{equation}
G^{\alpha \beta }=KT^{\alpha \beta }  \label{campo5}
\end{equation}
since the energy momentum tensor $T^{\alpha \beta }$ has now a fifth column
and row containing the electric current

\begin{equation}
T^{\alpha 5}=T^{5\alpha }=\rho _{0}v^{\alpha }v^{5}=\left[ 
\begin{array}{l}
\rho _{0}v^{1}\frac{q}{m} \\ 
\rho _{0}v^{2}\frac{q}{m} \\ 
\rho _{0}v^{3}\frac{q}{m} \\ 
\rho _{0}c\frac{q}{m} \\ 
\rho _{0}\frac{q^{2}}{m^{2}}
\end{array}
\right] =J^{\alpha }  \label{cor5}
\end{equation}
and $K$ is the coupling constant. The homogeneous equations can be derived,
as suggested by Kaluza, from the identity

\begin{equation}
F^{\lambda \alpha ,\beta }+F^{\alpha \beta ,\lambda }+F^{\beta \lambda
,\alpha }=0  \label{idfar}
\end{equation}
which arises from applying the cylindrical condition (eq. \ref{cil5}), to
the identity

\begin{equation}
\left( \Gamma _{\beta \lambda }^{\alpha }+\Gamma _{\lambda \alpha }^{\beta
}+\Gamma _{\alpha \beta }^{\lambda }\right) _{,\mu }=\Gamma _{\mu \lambda
}^{\alpha ,\beta }+\Gamma _{\mu \alpha }^{\beta ,\lambda }+\Gamma _{\mu
\beta }^{\lambda ,\alpha }  \label{idchris}
\end{equation}
taking $\mu =5$ .

In cartesian coordinates this mechanism yields the complete set of Maxwell
equations. However, if one works in generalized coordinates, Eq. (\ref
{idchris}) does not correspond to the correct sourceless electromagnetic
equations, a fact which will be shown in the next section. A strong
objection can be primarily made to Eq. (\ref{idchris}). Christoffel
symbols by themselves are not tensors and, although some combinations of
them are, Eq. (\ref{idchris}) is not tensorial, in contrast with Eq. 
\ref{campo5} which has tensors in both sides.

Summarizing, the theory has its weak point in the way sourceless equations
are obtained. On the other hand, the mechanism by which inhomogeneous
equations are derived, Einstein's field equation, is irrefutable and far
more elegant.

\section{\bf Five dimensions and generalized coordinates}

In this section we give the necessary arguments to circumvent the objection
raised in the preceding section regarding the derivation of Maxwell's
sourceless equations following Kaluza's procedure. To do so and for the sake
of simplicity we shall take as an example a Minkowski space with spherical
symmetry. The covariant and contravariant metric tensors are thus given by,

\begin{equation}
g_{\mu \nu }=\left[ 
\begin{array}{lllll}
1 & 0 & 0 & 0 & A_{1} \\ 
0 & r^{2} & 0 & 0 & r^{2}A_{2} \\ 
0 & 0 & r^{2}sin^{2}\theta & 0 & r^{2}sin^{2}\theta A_{3} \\ 
0 & 0 & 0 & -1 & -\frac{1}{c}\phi \\ 
A_{1} & r^{2}A_{2} & r^{2}sin^{2}\theta A_{3} & -\frac{1}{c}\phi & 1
\end{array}
\right]  \label{metesf5con}
\end{equation}

\begin{equation}
g^{\mu \nu }=\left[ 
\begin{array}{lllll}
1 & 0 & 0 & 0 & -A_{1} \\ 
0 & \frac{1}{r^{2}} & 0 & 0 & -A_{2} \\ 
0 & 0 & \frac{1}{r^{2}sin^{2}\theta } & 0 & -A_{3} \\ 
0 & 0 & 0 & -1 & \frac{1}{c}\phi \\ 
-A_{1} & -A_{2} & -A_{3} & \frac{1}{c}\phi & 1
\end{array}
\right]  \label{metesf5cov}
\end{equation}

In this metric which is similar to the one used in Kaluza's article, the two
metric tensors are inverse discarding quadratic terms. This means that (\ref
{metesf5con}) is the inverse of (\ref{metesf5cov}) only if second order terms
are eliminated. This approximation seems to be correct since it was also
used in Kaluza's article with a cartesian metric. Geometric coefficients
must be added to the components of the electromagnetic potential which
simply implies working in physical coordinates in order to obtain geometric
terms and coefficients in the equations. If $A_{\nu }^{ten}$ are tensorial
components of a vector, its physical components are computed as follows \ \ 
\begin{equation}
A_{\nu }^{phys}=\sqrt{g_{_{_{\nu \nu }}}}A_{\nu }^{ten}  \label{tenphys}
\end{equation}

It must be pointed out that, with these metric coefficients $g^{ij},$,
permutations of indices in the Christoffel symbols of Eq. (\ref{idchris})
do not {\em only} yield a change in sign as with a cartesian metric in which 
$g^{ij}=\delta ^{ij}$ with $i$ and $j$ running from 1 to 3, but also metric
coefficients relate these permutations . Moreover, the equation

\begin{equation}
\Gamma _{\beta 5}^{\alpha }=-\frac{g^{\alpha \alpha }}{g^{\beta \beta }}%
\Gamma _{\alpha 5}^{\beta }  \label{relchris}
\end{equation}
gives a relation between permutations of indices in Christoffel symbols
which makes equation (\ref{idchris}) dependent on the value given to the
indices unless the metric is cartesian, resulting in more than four
incorrect equations since each permutation of indices leads to an equation
with different coefficients. To clarify these statements let us take, for
example, the third component of the following equation

\begin{equation}
\varepsilon _{\beta \gamma }^{\nu }\left( \frac{\partial E_{\beta }}{%
\partial x^{\gamma }}-\frac{\partial E_{\gamma }}{\partial x^{\beta }}%
\right) =-\frac{1}{c}\frac{\partial B^{\nu }}{\partial t}  \label{maxrote}
\end{equation}
In the space we are working in, both sides of this equation in terms of
electromagnetic potentials should read

\begin{equation}
-\frac{2}{rsin\theta }\frac{\partial A_{2}}{\partial t}-\frac{1}{sin\theta }%
\frac{\partial ^{2}A_{2}}{\partial t\partial r}-\frac{1}{r^{2}sin\theta }%
\frac{\partial ^{2}A_{1}}{\partial t\partial \theta }  \label{maxbien}
\end{equation}
while Eq. (\ref{idchris}), taking $\alpha =4,$ $\beta =2,$ $\lambda =1$
and $\mu =5$ turns out to be as follows

\begin{equation}
-r\frac{\partial A_{2}}{\partial t}-\frac{1}{cr^{3}}\frac{\partial \phi }{%
\partial \theta }+\frac{\partial ^{2}A_{1}}{\partial \theta \partial t}+%
\frac{1}{2}\left( 1-r^{2}\right) \frac{\partial ^{2}A_{2}}{\partial
r\partial t}+\frac{1}{2c}\left( 1+\frac{1}{r^{2}}\right) \frac{\partial
^{2}\phi }{\partial r\partial \theta }=0  \label{maxmal}
\end{equation}
which can be written as

\begin{equation}
\frac{1}{2r}\frac{\partial ^{2}A_{2}}{\partial r\partial t}-\frac{1}{cr^{4}}%
\frac{\partial \phi }{\partial \theta }+\frac{1}{2cr}\left( 1+\frac{1}{r^{2}}%
\right) \frac{\partial ^{2}\phi }{\partial r\partial \theta }=\frac{\partial
A_{2}}{\partial t}+\frac{r}{2}\frac{\partial ^{2}A_{2}}{\partial r\partial t}%
-\frac{1}{2r}\frac{\partial ^{2}A_{1}}{\partial \theta \partial t}
\label{maxmal1}
\end{equation}
Eq. (\ref{maxmal1}) is obviously wrong. In Eq. (\ref{maxrote}), no
derivatives of the scalar potential should appear since its right hand side
is a time derivative of ${\bf B}$ which is only in terms of the vector
potential. The left hand side contains the rotational of the gradient of $%
\phi $ which is equal to zero. Similar results are obtained by changing the
indices for the rest of the equations. These equations are derived in
Appendix A.

On the other hand, the inhomogeneous Maxwell equations are correctly
recovered using Einstein's field equation as was mentioned. This induces us
to reformulate the theory in order to be able to obtain electromagnetic
equations, within Kaluza's formalism, based {\em only} in Einstein's field
equation without having to invoke the cylindrical condition to work on the
geometrical identity given by Eq. (\ref{idchris}).

\section{\bf Kaluza's formalism in 6 dimensions.}

The alternative here presented for Kaluza's formalism is based on a 6x6
metric. This space-time can thus be considered as an extension of Kaluza's
case. To proceed we define the following metric tensors with components
accounting for spherical symmetry namely,

\begin{equation}
g_{\mu \nu }=\left[ 
\begin{array}{llllll}
1 & 0 & 0 & 0 & A_{1} & Z_{1} \\ 
0 & r^{2} & 0 & 0 & r^{2}A_{2} & r^{2}Z_{2} \\ 
0 & 0 & r^{2}sin^{2}\theta & 0 & r^{2}sin^{2}\theta A_{3} & 
r^{2}sin^{2}\theta Z_{3} \\ 
0 & 0 & 0 & -1 & -\frac{1}{c}\phi & -\frac{1}{c}\eta \\ 
A_{1} & r^{2}A_{2} & r^{2}sin^{2}\theta A_{3} & \frac{1}{c}\phi & 1 & g_{56}
\\ 
Z_{1} & r^{2}Z_{2} & r^{2}sin^{2}\theta Z_{3} & -\frac{1}{c}\eta & g_{65} & 1
\end{array}
\right]  \label{metesf6cov}
\end{equation}

\begin{equation}
g^{\mu \nu }=\left[ 
\begin{array}{llllll}
1 & 0 & 0 & 0 & -A_{1} & -Z_{1} \\ 
0 & \frac{1}{r^{2}} & 0 & 0 & -A_{2} & -Z_{2} \\ 
0 & 0 & \frac{1}{r^{2}sin^{2}\theta } & 0 & -A_{3} & -Z_{3} \\ 
0 & 0 & 0 & -1 & \frac{1}{c}\phi & \frac{1}{c}\eta \\ 
-A_{1} & -A_{2} & -A_{3} & \frac{1}{c}\phi & 1 & g_{56} \\ 
-Z_{1} & -Z_{2} & -Z_{3} & \frac{1}{c}\eta & g_{65} & 1
\end{array}
\right]  \label{metesf6con}
\end{equation}

The structure of these tensors is proposed to maintain symmetry between the
fifth and sixth dimensions. The quantities $Z_{n}$ and $\eta $ are left
unspecified for the time being but will be interpreted later when the role
that each one plays in the equations becomes clear. Also $g_{56}=g_{65}$ is
left unsettled but it has to be proposed as time independent in order to
recover the conventional definitions for the fields (see Appendix B). The
position and velocity vectors are proposed, just following Kaluza, as follows

\begin{equation}
x^{\nu }=\left[ 
\begin{array}{c}
r \\ 
\theta \\ 
\varphi \\ 
ct \\ 
x^{5} \\ 
x^{6}
\end{array}
\right]  \label{pos6esf}
\end{equation}

\begin{equation}
v^{\nu }=\left[ 
\begin{array}{c}
v^{1} \\ 
v^{2} \\ 
v^{3} \\ 
c \\ 
\frac{q}{m} \\ 
\frac{\partial x^{6}}{\partial t}
\end{array}
\right]  \label{vel6esf}
\end{equation}
The equation of motion to be considered in this formalism is that of a
particle under the influence of a generalized Lorentz's force namely,

\begin{equation}
\frac{d^{2}x^{\alpha }}{dt^{2}}=\frac{q}{m}\left[ \varepsilon _{\beta \gamma
}^{\alpha }\frac{\partial x^{\beta }}{\partial t}B^{\gamma }+E^{\alpha
}\right] +\frac{g}{m}\left[ \varepsilon _{\beta \gamma }^{\alpha }\frac{%
\partial x^{\beta }}{\partial t}E^{\gamma }-B^{\alpha }\right]
\label{lor5gengen}
\end{equation}
This space is a more general one and has enough room to work in a
generalized scheme as will be soon shown. Here $g$ stands for the magnetic
charge. The Christoffel symbols that appear in the equation of a geodesic
(Eq.\ref{geod}) in this space are to be compared with the coefficients of
velocity components and charge in Eq. (\ref{lor5gengen}).

On the other hand, with the proposed metric, the  Christoffel symbols can also be
computed from definition (\ref{chrisgen}). The symbols to be proportional to
the electromagnetic field components are shown here as they arise directly
by introducing the elements of the metric tensors (\ref{metesf6con}) and (\ref
{metesf6cov}) in definition (\ref{chrisgen}). These symbols, in terms of
potentials become

\begin{equation}
\Gamma _{\beta 5}^{\alpha }=g^{\alpha \alpha }\left( A_{\beta ,\alpha
}-A_{\alpha ,\beta }+Z_{\alpha }g_{56,\beta }\right)  \label{chris5esf}
\end{equation}

\begin{equation}
\Gamma _{\beta 6}^{\alpha }=g^{\alpha \alpha }\left( Z_{\beta ,\alpha
}-Z_{\alpha ,\beta }+A_{\alpha }g_{56,\beta }\right)  \label{chris6esf}
\end{equation}
Comparing the symbols obtained from the geodesic and Eq. (\ref
{lor5gengen}) with the ones obtained directly from their definition,
expressions for electric and magnetic fields arise. Since we now have two
sets of Christoffel symbols, there are two expressions for each field (see
Appendix B).

\begin{equation}
E^{\nu }=-\frac{\partial \phi }{\partial x^{\nu }}-\frac{\partial A_{\nu }}{%
\partial t}  \label{defEconv}
\end{equation}

\begin{equation}
E^{\nu }=\varepsilon _{\beta \gamma }^{\nu }\left( \frac{\partial Z_{\beta }%
}{\partial x^{\gamma }}-\frac{\partial Z_{\gamma }}{\partial x^{\beta }}%
\right) +M^{\nu }  \label{defEnuev}
\end{equation}

\begin{equation}
B^{\nu }=-\frac{\partial \eta }{\partial x^{\nu }}-\frac{\partial Z_{\nu }}{%
\partial t}  \label{defBnuev}
\end{equation}

\begin{equation}
B^{\nu }=\varepsilon _{\beta \gamma }^{\nu }\left( \frac{\partial A_{\beta }%
}{\partial x^{\gamma }}-\frac{\partial A_{\gamma }}{\partial x^{\beta }}%
\right) +Q^{\nu }  \label{defBconv}
\end{equation}
where the vectors $M^{\nu }$ and $Q^{\nu }$ are defined as

\begin{equation}
M^\nu =\left[ 
\begin{array}{l}
Z_2\frac{\partial g_{56}}{\partial x^3} \\ 
Z_3\frac{\partial g_{56}}{\partial x^1} \\ 
Z_1\frac{\partial g_{56}}{\partial x^2}
\end{array}
\right] \;\;\qquad ,\qquad Q^\nu =\left[ 
\begin{array}{l}
A_2\frac{\partial g_{56}}{\partial x^3} \\ 
A_3\frac{\partial g_{56}}{\partial x^1} \\ 
A_1\frac{\partial g_{56}}{\partial x^2}
\end{array}
\right]  \label{defMyQ}
\end{equation}
These vectors are irrotational and since curls are solenoidal vectors,
expressions (\ref{defEnuev}) and (\ref{defBconv}) are in accordance with a
generalized Helmolhtz theorem for tensors\cite{Kobe}. The duality in
Faraday's tensor has been previously exhibited \cite{Kobe} as based in this
theorem but dual expressions for the fields are now introduced. Vectorial
fields can always be decomposed as a sum of a solenoidal vector field and an
irrotational one which validates expressions (\ref{defEnuev}) and (\ref
{defBconv}). Thus two sets of symmetrical expressions for the fields are
obtained one of these sets being the expressions of the decomposition
mentioned in the previous line. The implications of Eqs. (\ref{defEconv})
to (\ref{defBconv}) will be discussed in section 5.

Within this framework, the complete set of Maxwell equations can be derived
(Appendix C). The inhomogeneous equations are obtained in the same way as in
the previous section. The new quantity $Q^{\nu }$ is irrotational and
doesn't affect the structure of $\varepsilon _{\beta \gamma }^{\nu }\left( 
\frac{\partial B_{\beta }}{\partial x^{\gamma }}-\frac{\partial B_{\gamma }}{%
\partial x^{\beta }}\right) $ while the other equation with sources is
identically obtained provided the same definition for $E$ is used. The
sourceless equations are obtained once again by means of Einstein's field
equation

\begin{equation}
R^{6\beta }=KT^{6\beta }  \label{campo6}
\end{equation}
since the energy momentum tensor has now one additional column and row

\begin{equation}
T^{6\beta }=T^{6\beta \alpha }=\rho _{0}v^{\beta }v^{6}=\left[ 
\begin{array}{l}
\rho _{0}v^{1}\frac{\partial x^{6}}{\partial t} \\ 
\rho _{0}v^{2}\frac{\partial x^{6}}{\partial t} \\ 
\rho _{0}v^{3}\frac{\partial x^{6}}{\partial t} \\ 
\rho _{0}c\frac{\partial x^{6}}{\partial t} \\ 
\rho _{0}\left( \frac{\partial x^{6}}{\partial t}\right) ^{2}
\end{array}
\right] =K^{\beta }  \label{momen6}
\end{equation}

In Appendix C it is proved that Eq. (\ref{campo6}) leads to the four
missing Maxwell equations with some extra terms that make the complete set
of equations symmetric. This new terms will be interpreted in next section.

Thus, the purpose of this section is accomplished. The mechanism for
obtaining both homogeneous and inhomogeneous equations is the same. This
makes the theory completely symmetric and the cylindrical condition not so
fundamental for the recovery of Maxwell's equations.

\section{\bf Implications of the sixth dimension}

In Kaluza's theory, the fifth dimension is associated with electric charge.
As electric charge generates curvature in space, it becomes intuitive that
the sixth dimension is a curvature source similar to it. The complete set of
equations (the equation of motion and Maxwell relations) as is obtained in
this formalism is given by

\begin{equation}
\frac{d^{2}x^{\alpha }}{dt^{2}}=\frac{q}{m}\left[ \varepsilon _{\beta \gamma
}^{\alpha }\frac{\partial x^{\beta }}{\partial t}B^{\gamma }+E^{\alpha
}\right] +\frac{g}{m}\left[ \varepsilon _{\beta \gamma }^{\alpha }\frac{%
\partial x^{\beta }}{\partial t}E^{\gamma }-B^{\alpha }\right]
\label{geodmonop}
\end{equation}

\begin{equation}
\nabla \times {\bf E}=-\frac{1}{c}\frac{\partial {\bf B}}{\partial t}-{\bf K}
\label{max1}
\end{equation}

\begin{equation}
\nabla \times {\bf B}=\frac{\partial {\bf E}}{\partial t}+\mu _{0}{\bf J}
\label{max2}
\end{equation}

\begin{equation}
\nabla \cdot {\bf E}=\frac{\rho }{\varepsilon _{0}}\frac{q}{m}
\label{max3}
\end{equation}

\begin{equation}
\nabla \cdot {\bf B}\;=\mu _0\rho \frac gm  \label{max4}
\end{equation}

This finally confirms that the sixth dimension is associated with a magnetic
charge $g$. Eqs. (\ref{max1}) and (\ref{max4}) show the presence of
magnetic charge and it's associated magnetic current ${\bf K}$. These
equations are completely symmetric as the new definitions (\ref{defEconv}) to 
(\ref{defBconv}) for electric and magnetic fields result in this formalism.
The new potentials $Z_\alpha $ and $\eta $ can be interpreted as
electromagnetic potentials. $Z_\alpha $ plays the role of an electric vector
potential and $\eta $ a scalar magnetic potential. Both appear in these
definitions but do not affect the behavior of charged particles since they
do not introduce extra effects in the equations. The magnetic charge and
current terms in the equations arise exclusively from the extra column in
the energy-momentum tensor $T^{6\beta }$. To recover the equations without
this extra terms, it is sufficient to make the magnetic charge equal to
zero, $\frac{\partial x^6}{\partial t}=0$, just as was done with the fifth
component of the position vector to recover the equations of motion for a
particle without electrical charge.

Also, from the combination of Eqs. (\ref{defEconv}) to (\ref{defBconv}) and
Maxwell's equations one can easily obtain four wave equations for
potentials, providing a Lorentz's gauge for the new four vector potential is
considered. In the procedure, continuity relations for electric and magnetic
charge are obtained.

The role of the vectors $M^\nu $ and $Q^\nu $ (Eq. \ref{defMyQ}) becomes
clear by taking a closer look to definitions (\ref{defEnuev}) and (\ref
{defBconv}). In these expressions, the fields are decomposed as a curl plus a
vector which is irrotational. In order to obtain both Maxwell equations that
feature charge density as a source, one has to calculate the divergence of
both fields. In the procedure, the divergence of the rotational term
vanishes since rotationals are solenoidal vectors leaving alone the
divergence of $M^\nu $ and $Q^\nu $ which should then be proportional to the
charge densities. If this vectors do not appear in the definitions, the
result would be solenoidal magnetic and {\em electric} fields meaning that
neither magnetic {\em nor electric} charges exist. The spatial derivatives
of the metric elements $g_{56}=g_{65}$ appear in $M^\nu $ and $Q^\nu $ which
makes its value fundamental since if it is proposed as a constant, these
vectors would be equal to zero with the consecuences mentioned before.

\section{\bf Conclusions}

Theodore Kaluza proposed many decades ago a way to unify electromagnetic and
gravitational theories. His formalism, although very useful and elegant,
shows a weak point in the way a pair of Maxwell's equations are obtained.
Moreover, this mechanism doesn't give correct results when working in
generalized coordinates which are necessary to get the correct physics of
most systems.

The alternative here presented is a similar formalism which consists in
working in a larger space-time. The new extra dimension makes it possible to
use Einstein's field equation twice and all the equations can be correctly
recovered. Even in the absence of this curvature source, the extra dimension
remains necessary.

Additional to the fact of leading to the complete and correct set of Maxwell
equations, this formalism allows the introduction of magnetic charge in a
very elegant and natural way. The metric and ensuing procedure are proposed
based only in a symmetrical way of treating electric and magnetic fields. As
a result of this, magnetic charge effects arise in the theory in the same
way electric effects do. Also, following the same procedure as in
conventional electromagnetism, wave equations for potentials in the presence
of magnetic charge, can be obtained without having to introduce any
singularity in space, as done in other works \cite{mdernkaluza}.

Some approximations have been made in order to verify the equations that
emerge from the classical treatment of electromagnetism. First of all, the
metric used neglects quadratic terms. These terms may introduce additional
measurable effects. In Kaluza's work, the basic hypothesis of a small
electric charge-mass ratio has to be considered. In the formalism here
presented a similar assumption has to be made about the magnetic charge-mass
ratio. This is of course a mere conjecture since no magnetic monopoles have
yet been detected.

The correct set of equations in generalized coordinates can also be obtained
using Bianchi's identities. We consider the formalism here presented more
elegant since charges and electromagnetic potentials are included in the
geometrical description of space-time and no external forces or fields have
to be introduced. Electromagnetic effects are consequences of space
configuration. Kaluza's theory has been found to be very useful for the
formulation of magnetohydrodynamics equations in the context of irreversible
process thermodynamics \cite{ti} and in the development of gauge theories 
\cite{Straumann} so that we consider fundamental the reformulation of the
theory in generalized coordinates to treat problems with different spatial
symmetries.

\section{\bf Appendix A}

Sourceless Maxwell's equations in vectorial notation read:

\begin{equation}
\nabla \times {\bf E}=-\frac{1}{c}\frac{\partial {\bf B}}{\partial t} 
\label{rote}
\end{equation}

\begin{equation}
\nabla \cdot {\bf B}\;=0  \label{divb}
\end{equation}
To write Eq. (\ref{rote})  in terms of potentials, first we express the vector
potential $A_\nu $\ and the gradient of the scalar potential $\phi $ in
physical coordinates:

\begin{equation}
A_{\nu }^{phys}=\left[ 
\begin{array}{c}
A_{1} \\ 
rA_{2} \\ 
rsin\theta A_{3}
\end{array}
\right]  \label{afis}
\end{equation}

\begin{equation}
\nabla \phi ^{phys}=\left[ 
\begin{array}{c}
\frac{\partial \phi }{\partial r} \\ 
\frac{1}{r}\frac{\partial \phi }{\partial \theta } \\ 
\frac{1}{rsin\theta }\frac{\partial \phi }{\partial \varphi }
\end{array}
\right]  \label{gradfis}
\end{equation}
The electric field is then

\begin{equation}
E^\nu {\bf =}\left[ 
\begin{array}{l}
-\frac{\partial A_1}{\partial t}-\frac{\partial \phi }{\partial r} \\ 
-\frac{\partial rA_2}{\partial t}-\frac 1r\frac{\partial \phi }{\partial
\theta } \\ 
-\frac{\partial rsin\theta A_3}{\partial t}-\frac{\partial \phi }{\partial
\varphi }
\end{array}
\right]  \label{eesf}
\end{equation}

To calculate the curl of ${\bf E}$, the following determinant has to be
expanded 
\begin{equation}
\nabla \times {\bf E=}\frac 1{r^2sin\theta }\left| 
\begin{array}{ccc}
\overrightarrow{r} & r\overrightarrow{\theta } & rsin\theta \overrightarrow{%
\varphi } \\ 
\frac \partial {\partial r} & \frac \partial {\partial \theta } & \frac
\partial {\partial \varphi } \\ 
E^1 & rE^2 & rsin\theta E^3
\end{array}
\right|  \label{detrote}
\end{equation}
The third component of the $\nabla \times {\bf E}$ is then

\begin{equation}
\frac{rsin\theta }{r^{2}sin\theta }\left[ \frac{\partial }{\partial r}\left(
r\left( -\frac{\partial rA_{2}}{\partial t}-\frac{1}{r}\frac{\partial \phi }{%
\partial \theta }\right) \right) -\frac{\partial }{\partial \theta }\left( -%
\frac{\partial A_{1}}{\partial t}-\frac{\partial \phi }{\partial r}\right)
\right] \overrightarrow{\varphi }  \label{rot3}
\end{equation}
After some algebra Eq. (\ref{rot3})  reduces to

\begin{equation}
-2\frac{\partial A_{2}}{\partial t}-r\frac{\partial ^{2}A_{2}}{\partial
t\partial r}+\frac{1}{r}\frac{\partial ^{2}A_{1}}{\partial t\partial \theta }
\label{rot31}
\end{equation}
Eq. (\ref{rot31})  has to be divided by $\sqrt{g_{33}}$ to return the expresion to
its tensorial components which finally results in Eq. (\ref{maxbien}).
Note that in equation \ref{rot31}  the scalar potential does not appear, which agrees
with the property of the gradient being irrotational.

On the other hand, the left hand side of equation \ref{idchris}, with $%
\alpha =4,$ $\beta =2,$ $\lambda =1$ and $\mu =5$ should be equal to zero
because of the cylindrical condition (Eq. \ref{cil5}):

\begin{equation}
\left( \Gamma _{21}^{4}+\Gamma _{14}^{2}+\Gamma _{42}^{1}\right) _{,5}=0 
\label{idcrhisin}
\end{equation}
Which means 
\begin{equation}
\Gamma _{51}^{4,2}+\Gamma _{54}^{2,1}+\Gamma _{52}^{1,4}=0  
\label{idchrisin1}
\end{equation}
The Christoffel symbols involved in Eq. (\ref{idcrhisin})  are computed as follows:

\begin{equation}
\Gamma _{51}^{4}=\frac{1}{2}g^{44}\left( \frac{\partial g_{54}}{\partial
x^{1}}+\frac{\partial g_{14}}{\partial x^{5}}-\frac{\partial g_{51}}{%
\partial x^{4}}\right) =\frac{1}{2c}\frac{\partial A_{1}}{\partial t}+\frac{1%
}{2c}\frac{\partial \phi }{\partial r}   \label{chris451}
\end{equation}

\begin{equation}
\Gamma _{54}^{2}=\frac{1}{2}g^{22}\left( \frac{\partial g_{52}}{\partial
x^{4}}+\frac{\partial g_{42}}{\partial x^{5}}-\frac{\partial g_{54}}{%
\partial x^{2}}\right) =\frac{1}{2cr^{2}}\frac{\partial r^{2}A_{2}}{\partial
t}+\frac{1}{2cr^{2}}\frac{\partial \phi }{\partial \theta } 
\label{chris254}
\end{equation}

\begin{equation}
\Gamma _{52}^{1}=\frac{1}{2}g^{11}\left( \frac{\partial g_{51}}{\partial
x^{2}}+\frac{\partial g_{21}}{\partial x^{5}}-\frac{\partial g_{52}}{%
\partial x^{1}}\right) =\frac{1}{2}\frac{\partial A_{1}}{\partial \theta }%
-r^{2}\frac{\partial A_{2}}{\partial r}-2rA_{2}  \label{chris152}
\end{equation}
Introducing the previous results in Eq. (\ref{idcrhisin})  we obtain an incorrect
expression for Eq. (\ref{rote})  , namely

\begin{equation}
-\frac rc\frac{\partial A_2}{\partial t}-\frac 1{2cr^3}\frac{\partial \phi }{%
\partial \theta }+\frac 1c\frac{\partial ^2A_1}{\partial \theta \partial t}%
+\frac 1{2c}\left( 1-r^2\right) \frac{\partial ^2A_2}{\partial r\partial t}%
+\frac 1{2c}\left( 1+\frac 1{r^2}\right) \frac{\partial ^2\phi }{\partial
r\partial \theta }=0
\end{equation}

\section{\bf Appendix B}

The equation for the geodesic is:

\begin{equation}
\frac{d^{2}x^{\alpha }}{ds^{2}}+\Gamma _{\mu \nu }^{\alpha }\frac{dx^{\mu }}{%
ds}\frac{dx^{\nu }}{ds}=0  \label{b1}
\end{equation}
In the six dimensional space proposed the first of the three equations in
the geodesic turns out as follows:

\begin{equation}
\frac{d^{2}r}{dt^{2}}+\Gamma _{5\nu }^{1}\frac{q}{m}\frac{dx^{\nu }}{ds}%
+\Gamma _{6\nu }^{1}\frac{g}{m}\frac{dx^{\nu }}{ds}=0 \label{b2}
\end{equation}
which after expanding the sums over $\nu $ reads

\begin{equation}
\begin{array}{l}
\frac{d^2r}{dt^2}+\Gamma _{51}^1\frac qm\frac{dx^1}{ds}+\Gamma _{52}^1\frac
qm\frac{dx^2}{ds}+\Gamma _{53}^1\frac qm\frac{dx^3}{ds}+\Gamma _{54}^1\frac
qm\frac{dx^4}{ds}+ \\ 
\;\;\;\Gamma _{61}^1\frac gm\frac{dx^1}{ds}+\Gamma _{62}^1\frac gm\frac{dx^2%
}{ds}+\Gamma _{63}^1\frac gm\frac{dx^3}{ds}+\Gamma _{64}^1\frac gm\frac{dx^4%
}{ds}=0
\end{array} \label{b3}
\end{equation}
On the other hand, the first component of the equation of motion under a
generalized Lorentz's (Eq. \ref{lor5gengen}) force is

\begin{equation}
\frac{d^2r}{dt^2}=\frac qm\frac{\partial x^2}{\partial t}B_3-\frac qm\frac{%
\partial x^3}{\partial t}B_2+\frac qmE_1+\frac gm\frac{\partial x^2}{%
\partial t}E_3-\frac gm\frac{\partial x^3}{\partial t}E_2-\frac gmB_1 
\label{b4}
\end{equation}
Then the relationship between the Christoffel symbols and the components of
the electromagnetic field should be, for this equation:

\begin{equation}
\Gamma _{51}^{1}=0  \label{b5}
\end{equation}

\begin{equation}
\Gamma _{52}^{1}=B_{3}  \label{b6}
\end{equation}

\begin{equation}
\Gamma _{53}^{1}=-B_{2}  \label{b7}
\end{equation}

\begin{equation}
\Gamma _{54}^{1}=\frac{1}{c}E_{1} \label{b8}
\end{equation}
and

\begin{equation}
\Gamma _{61}^{1}=0  \label{b9}
\end{equation}

\begin{equation}
\Gamma _{62}^{1}=E_{3}  \label{b10}
\end{equation}

\begin{equation}
\Gamma _{63}^{1}=-E_{2}  \label{b11}
\end{equation}

\begin{equation}
\Gamma _{64}^{1}=-\frac{1}{c}B_{1}  \label{b12}
\end{equation}
Calculating the symbols $\Gamma _{52}^{1}$\ and $\Gamma _{64}^{1}$\ directly
from the definition \ref{chrisgen} we have

\begin{equation}
\Gamma _{52}^{1}=\frac{1}{2}g^{11}\left( \frac{\partial g_{51}}{\partial
x^{2}}-\frac{\partial g_{52}}{\partial x^{1}}\right) +\frac{1}{2}g^{16}\frac{%
\partial g_{56}}{\partial x^{2}}=\frac{1}{2}\left( \frac{\partial A_{1}}{%
\partial x^{2}}-\frac{\partial A_{2}}{\partial x^{1}}\right) +\frac{1}{2}%
Z_{1}\frac{\partial g_{56}}{\partial x^{2}}=B_{3}  \label{b13}
\end{equation}

\begin{equation}
\Gamma _{64}^{1}=\frac{1}{2}g^{11}\left( \frac{\partial g_{61}}{\partial
x^{4}}-\frac{\partial g_{64}}{\partial x^{1}}\right) +\frac{1}{2}g^{15}\frac{%
\partial g_{56}}{\partial x^{4}}=\frac{1}{2}\left( \frac{\partial Z_{1}}{%
\partial x^{4}}+\frac{1}{c}\frac{\partial \eta }{\partial x^{1}}\right) =-%
\frac{1}{c}B_{1}  \label{b14}
\end{equation}

In equation (\ref{b14})  the quantity $g_{56}$ is supposed as time independent in
order to recover the conventional definition of the magnetic field.
Eqs (\ref{b13})    and (\ref{b14})  exhibit how the magnetic field can be represented in
two different ways (Eqs. (\ref{defBconv}) and (\ref{defBnuev})). Similar
results, now verifying equations (\ref{defEconv}) and (\ref{defEnuev}), are
obtained by computing $\Gamma _{54}^{1}$ and $\Gamma _{62}^{1}$ from
definition (\ref{chrisgen}) namely,

\begin{equation}
\Gamma _{54}^{1}=\frac{1}{2}g^{11}\left( \frac{\partial g_{51}}{\partial
x^{4}}-\frac{\partial g_{54}}{\partial x^{1}}\right) +\frac{1}{2}g^{16}\frac{%
\partial g_{56}}{\partial x^{4}}=\frac{1}{2}\left( \frac{\partial A_{1}}{%
\partial x^{4}}+\frac{1}{c}\frac{\partial \phi }{\partial x^{1}}\right) =%
\frac{1}{c}E_{1}  \label{b15}
\end{equation}

\begin{equation}
\Gamma _{62}^{1}=\frac{1}{2}g^{11}\left( \frac{\partial g_{61}}{\partial
x^{2}}-\frac{\partial g_{62}}{\partial x^{1}}\right) +\frac{1}{2}g^{15}\frac{%
\partial g_{65}}{\partial x^{2}}=\frac{1}{2}\left( \frac{\partial Z_{1}}{%
\partial x^{2}}-\frac{\partial Z_{2}}{\partial x^{1}}\right) +\frac{1}{2}%
A_{1}\frac{\partial g_{65}}{\partial x^{2}}=E_{3}  \label{b16}
\end{equation}

\section{\bf Appendix C}

In this appendix it is shown how the complete set of correct Maxwell
equations is obtained only by means of Einstein's equation. In the first
part, equations with sources are obtained. This procedure can be carried out
in a 5x5 space-time giving the same results, since the Christoffel symbols
required for the following operations are the ones with one index equal to 5.

Einstein's field equation relates Ricci's tensor with the mass-energy
tensor. If we assume the curvature scalar to be cero, the field equation
turns out as follows:

\begin{equation}
R_{\mu \nu }=KT_{\mu \nu }  \label{campob}
\end{equation}
where Ricci's tensor is defined as \cite{Stephani}

\begin{equation}
R_{\mu \nu }=\Gamma _{\mu \nu ,n}^n+\Gamma _{mn}^n\Gamma _{\mu \nu
}^m-\Gamma _{\mu m}^n\Gamma _{\nu n}^m  \label{riccib}
\end{equation}
In order to obtain the inhomogeneous equations, let's take $\mu =5$ and let $%
\nu $ run from 1 to 4.

\begin{equation}
\begin{array}{l}
R_{51}=\frac{\cot \theta }rA_2+\frac 12\frac{\partial ^2A_1}{\partial t^2}%
+\frac 1r\frac{\partial A_3}{\partial \varphi }-\frac{\csc ^2\theta }{r^2}%
\frac{\partial ^2A_1}{\partial \varphi ^2}-\frac{\cot \theta }{2r^2}\frac{%
\partial A_1}{\partial \theta }+\frac 1r\frac{\partial A_2}{\partial \theta }%
-\frac 1{2r^2}\frac{\partial ^2A_1}{\partial \theta ^2}+ \\ 
\;\;\;\;\;\;\;\frac{\cot \theta }2\frac{\partial A_2}{\partial r}+\frac 1{2c}%
\frac{\partial ^2\phi }{\partial r\partial t}+\frac 12\frac{\partial ^2A_3}{%
\partial r\partial \varphi }+\frac 12\frac{\partial ^2A_2}{\partial
r\partial \theta }
\end{array}
\label{c3}
\end{equation}

\begin{equation}
\begin{array}{l}
R_{52}=-A_2+\frac{r^2}2\frac{\partial ^2A_2}{\partial t^2}+\cot \theta \frac{%
\partial A_3}{\partial \varphi }-\frac{\csc ^2\theta }2\frac{\partial ^2A_2}{%
\partial \varphi ^2}+\frac 1{2c}\frac{\partial ^2\phi }{\partial \theta
\partial t}+ \\ 
\;\;\;\;\;\;\;\frac 12\frac{\partial ^2A_3}{\partial \theta \partial \varphi 
}-2r\frac{\partial A_2}{\partial r}+\frac 12\frac{\partial ^2A_1}{\partial
r\partial \theta }-\frac{r^2}2\frac{\partial ^2A_2}{\partial r^2}
\end{array}
\label{c4}
\end{equation}

\begin{equation}
\begin{array}{l}
R_{53}=\frac{r^2}2\sin ^2\theta \frac{\partial ^2A_3}{\partial t^2}-\frac{%
\cot \theta }2\frac{\partial A_2}{\partial \varphi }+\frac 1{2c}\frac{%
\partial ^2\phi }{\partial \varphi \partial t}-\frac 32\cos \theta \sin
\theta \frac{\partial A_3}{\partial \theta }+\frac 12\frac{\partial ^2A_2}{%
\partial \theta \partial \varphi }- \\ 
\;\;\;\;\;\;\;\frac 12\sin ^2\theta \frac{\partial ^2A_3}{\partial \theta ^2}%
-2r\sin ^2\theta \frac{\partial A_3}{\partial r}+\frac 12\frac{\partial ^2A_1%
}{\partial r\partial \varphi }-\frac{r^2}2\sin ^2\theta \frac{\partial ^2A_3%
}{\partial r^2}
\end{array}
\label{c5}
\end{equation}

\begin{equation}
\begin{array}{l}
R_{54}=\frac 1{2r}\frac{\partial A_1}{\partial t}+\frac{\cot \theta }2\frac{%
\partial A_2}{\partial t}+\frac 12\frac{\partial ^2A_3}{\partial \varphi
\partial t}+\frac{\csc ^2\theta }{2cr^2}\frac{\partial ^2\phi }{\partial
\varphi ^2}+\frac{\cot \theta }{2cr^2}\frac{\partial \phi }{\partial \theta }%
+\frac 12\frac{\partial ^2A_2}{\partial \theta \partial t}+ \\ 
\;\;\;\;\;\;\;\frac 1{2cr^2}\frac{\partial ^2\phi }{\partial \theta ^2}%
+\frac 1{2cr}\frac{\partial \phi }{\partial r}+\frac 12\frac{\partial ^2A_1}{%
\partial r\partial t}+\frac 1{2c}\frac{\partial ^2\phi }{\partial r^2}
\end{array}
\label{c6}
\end{equation}

If each one of these equations is equaled with its corresponding element of
the energy-momentum tensor $T_{\mu 5}$ (Eq. \ref{cor5}), Maxwell equations
with sources are correctly recovered.

To obtain the homogeneous equations, the sixth row of Ricci's tensor has to
be calculated. The procedure is exactly the same as the one to obtain the
inhomogeneous equations and the results are:

\begin{equation}
\begin{array}{l}
R_{61}=\frac{\cot \theta }rZ_2+\frac 12\frac{\partial ^2Z_1}{\partial t^2}%
+\frac 1r\frac{\partial Z_3}{\partial \varphi }-\frac{\csc ^2\theta }{r^2}%
\frac{\partial ^2Z_1}{\partial \varphi ^2}-\frac{\cot \theta }{2r^2}\frac{%
\partial Z_1}{\partial \theta }+\frac 1r\frac{\partial Z_2}{\partial \theta }%
-\frac 1{2r^2}\frac{\partial ^2Z_1}{\partial \theta ^2}+ \\ 
\;\;\;\;\;\;\;\frac{\cot \theta }2\frac{\partial Z_2}{\partial r}+\frac 1{2c}%
\frac{\partial ^2\eta }{\partial r\partial t}+\frac 12\frac{\partial ^2Z_3}{%
\partial r\partial \varphi }+\frac 12\frac{\partial ^2Z_2}{\partial
r\partial \theta }
\end{array}
\label{c7}
\end{equation}

\begin{equation}
\begin{array}{l}
R_{62}=-Z_2+\frac{r^2}2\frac{\partial _2^2Z}{\partial t^2}+\cot \theta \frac{%
\partial Z_3}{\partial \varphi }-\frac{\csc ^2\theta }2\frac{\partial ^2Z_2}{%
\partial \varphi ^2}+\frac 1{2c}\frac{\partial ^2\eta }{\partial \theta
\partial t}+ \\ 
\;\;\;\;\;\;\;\frac 12\frac{\partial _3^2Z}{\partial \theta \partial \varphi 
}-2r\frac{\partial Z_2}{\partial r}+\frac 12\frac{\partial _1^2Z}{\partial
r\partial \theta }-\frac{r^2}2\frac{\partial _2^2Z}{\partial r^2}
\end{array}
\label{c8}
\end{equation}

\begin{equation}
\begin{array}{l}
R_{63}=\frac{r^2}2\sin ^2\theta \frac{\partial ^2Z_3}{\partial t^2}-\frac{%
\cot \theta }2\frac{\partial Z_2}{\partial \varphi }+\frac 1{2c}\frac{%
\partial ^2\eta }{\partial \varphi \partial t}-\frac 32\cos \theta \sin
\theta \frac{\partial Z_3}{\partial \theta }+\frac 12\frac{\partial ^2Z_2}{%
\partial \theta \partial \varphi }- \\ 
\;\;\;\;\;\;\;\frac 12\sin ^2\theta \frac{\partial ^2Z_3}{\partial \theta ^2}%
-2r\sin ^2\theta \frac{\partial Z_3}{\partial r}+\frac 12\frac{\partial ^2Z_1%
}{\partial r\partial \varphi }-\frac{r^2}2\sin ^2\theta \frac{\partial ^2Z_3%
}{\partial r^2}
\end{array}
\label{c9}
\end{equation}

\begin{equation}
\begin{array}{l}
R_{64}=\frac 1{2r}\frac{\partial Z_1}{\partial t}+\frac{\cot \theta }2\frac{%
\partial Z_2}{\partial t}+\frac 12\frac{\partial ^2Z_3}{\partial \varphi
\partial t}+\frac{\csc ^2\theta }{2cr^2}\frac{\partial ^2\eta }{\partial
\varphi ^2}+\frac{\cot \theta }{2cr^2}\frac{\partial \eta }{\partial \theta }%
+\frac 12\frac{\partial ^2Z_2}{\partial \theta \partial t}+ \\ 
\;\;\;\;\;\;\;\frac 1{2cr^2}\frac{\partial ^2\eta }{\partial \theta ^2}%
+\frac 1{2cr}\frac{\partial \eta }{\partial r}+\frac 12\frac{\partial ^2Z_1}{%
\partial r\partial t}+\frac 1{2c}\frac{\partial ^2\eta }{\partial r^2}
\end{array}
\label{c10}
\end{equation}

If we introduce expressions (\ref{defEnuev}) and (\ref{defBnuev}) as definitions
for ${\bf E}$ and ${\bf B}$ respectively, and set each of the previous
equations ( \ref{c7}) to (\ref{c10}) equal to its corresponding element in the
energy-momentum tensor $T_{\mu 6}$ Eqs. (\ref{max1}) and (\ref{max4}) are
obtained. For example, to prove equation  \ref{c9}  corresponds to the third
component of equation (\ref{max1}) we will calculate the third component or
the curl of the electric field using the alternative expression in Eq. 
(\ref{defEnuev}). First, the vector potential $Z_\beta $ has to be expressed
in its physical components as follows

\begin{equation}
Z_{\nu }^{phys}=\left[ 
\begin{array}{c}
Z_{1} \\ 
rZ_{2} \\ 
rsin\theta Z_{3}
\end{array}
\right]  \label{c11}
\end{equation}
The curl of $Z_{\nu }^{phys}$ is calculated by expanding following the
determinat

\begin{equation}
\begin{array}{l}
\nabla \times {\bf Z=}\frac 1{r^2sin\theta }\left| 
\begin{array}{ccc}
\overrightarrow{r} & r\overrightarrow{\theta } & rsin\theta \overrightarrow{%
\varphi } \\ 
\frac \partial {\partial r} & \frac \partial {\partial \theta } & \frac
\partial {\partial \varphi } \\ 
Z_1 & r\left( rZ_2\right) & rsin\theta \left( rsin\theta Z_3\right)
\end{array}
\right| =\frac 1{r^2sin\theta }\left[ 
\begin{array}{c}
r^2\frac{\partial sin^2\theta Z_3}{\partial \theta }-r^2\frac{\partial Z_2}{%
\partial \varphi } \\ 
r\left( \frac{\partial Z_1}{\partial \varphi }-sin^2\theta \frac{\partial
r^2Z_3}{\partial r}\right) \\ 
rsin\theta \left( \frac{\partial r^2Z_2}{\partial r}-\frac{\partial Z_1}{%
\partial \theta }\right)
\end{array}
\right] \\ 
\;\;\;\;\;\;\;\;\;\;\;\;\;\;\;\;\;\;\;\;=\left[ 
\begin{array}{c}
sin\theta \frac{\partial Z_3}{\partial \theta }+2\cos \theta Z_3-\csc \theta 
\frac{\partial Z_2}{\partial \varphi } \\ 
\frac 1{r\sin \theta }\frac{\partial Z_1}{\partial \varphi }-r\sin \theta 
\frac{\partial Z_3}{\partial r}-2sin\theta Z_3 \\ 
r\frac{\partial Z_2}{\partial r}+2Z_2-\frac 1r\frac{\partial Z_1}{\partial
\theta }
\end{array}
\right]
\end{array}
\label{c12}
\end{equation}

Since $M^\nu $ is irrotational, $\nabla \times {\bf E=}\nabla \times \left(
\nabla \times {\bf Z}\right) $ which is calculated as follows

\begin{equation}
\nabla \times {\bf E=}\frac 1{r^2sin\theta }\left| 
\begin{array}{ccc}
\overrightarrow{r} & r\overrightarrow{\theta } & rsin\theta \overrightarrow{%
\varphi } \\ 
\frac \partial {\partial r} & \frac \partial {\partial \theta } & \frac
\partial {\partial \varphi } \\ 
\left[ \nabla \times {\bf Z}\right] ^1 & r\left[ \nabla \times {\bf Z}%
\right] ^2 & rsin\theta \left[ \nabla \times {\bf Z}\right] ^3
\end{array}
\right|  \label{c13}
\end{equation}
The third component of the last equation reads

\begin{equation}
\frac{1}{r}\left[ \frac{\partial }{\partial r}\left( r\left( \frac{1}{r\sin
\theta }\frac{\partial Z_{1}}{\partial \varphi }-r\sin \theta \frac{\partial
Z_{3}}{\partial r}-2sin\theta Z_{3}\right) \right) -\frac{\partial }{%
\partial \theta }\left( sin\theta \frac{\partial Z_{3}}{\partial \theta }%
+2\cos \theta Z_{3}-\csc \theta \frac{\partial Z_{2}}{\partial \varphi }%
\right) \right]  \label{c14}
\end{equation}
which after some algebraic manipulation turns out as follows

\begin{equation}
\frac{1}{r\sin \theta }\frac{\partial ^{2}Z_{1}}{\partial r\partial \varphi }%
-r\sin \theta \frac{\partial ^{2}Z_{3}}{\partial r^{2}}-4\sin \theta \frac{%
\partial Z_{3}}{\partial r}-\frac{sin\theta }{r}\frac{\partial ^{2}Z_{3}}{%
\partial \theta ^{2}}-\frac{3\cos \theta }{r}\frac{\partial Z_{3}}{\partial
\theta }-\frac{\cos \theta }{r\sin ^{2}\theta }\frac{\partial Z_{2}}{%
\partial \varphi }+\frac{\csc \theta }{r}\frac{\partial ^{2}Z_{2}}{\partial
\theta \partial \varphi }  \label{c15}
\end{equation}

Expression  (\ref{c15})  is the third physical component of the curl of the electric
field. To obtain the third component of Eq. (\ref{max1}), also $-\frac{1}{%
c}\frac{\partial {\bf B}}{\partial t}-{\bf K}$ has to be in physical
components which implies multiplying by $r\sin \theta $. Then, to make it
clear that equation (\ref{c15})  corresponds to (\ref{max1}) it has to be expressed in
the following way:

\begin{equation}
\frac 1{r\sin \theta }\left( \frac{\partial ^2Z_1}{\partial r\partial
\varphi }-r^2\sin ^2\theta \frac{\partial ^2Z_3}{\partial r^2}-4r\sin
^2\theta \frac{\partial Z_3}{\partial r}-sin^2\theta \frac{\partial ^2Z_3}{%
\partial \theta ^2}-\frac 32\sin 2\theta \frac{\partial Z_3}{\partial \theta 
}-\frac{\cos \theta }{\sin \theta }\frac{\partial Z_2}{\partial \varphi }+%
\frac{\partial ^2Z_2}{\partial \theta \partial \varphi }\right) 
\label{c16}
\end{equation}
This finally proves that Eqs. (\ref{c7})   to (\ref{c10})    correspond to the missing
Maxwell's equations (\ref{max1}) and (\ref{max4}). \bigskip \ These equations
include magnetic sources. It is important to point out, once again, that to
obtain Maxwell's equations without magnetic sources it is sufficient to make 
$\frac{\partial x^6}{\partial t}=0$ which implies $T_{\mu 6}=0$.


\begin{thebibliography}{1}
\bibitem[1]{Kaluza}  Th. Kaluza. Sitzungsberg. D. Preiss. Akademie D.
Wissenschaften, Physik.-Mathemat. Klasse, 966-972 (1921).

\bibitem[2]{Straumann}  L O'Raifertaigh, N. Straumann; Rev. Mod. Phys. 72,1
(2000) and references cited therein.

\bibitem[3]{Mesiner}  C. Mesiner W., K.S Thorne, J.A
Wheeler.''Gravitation'', Ed. W.H.Freeman \& Co., 1973.

\bibitem[4]{Kobe}  D.H. Kobe. Am. J. Phys. 52, April 1984.

\bibitem[5]{mdernkaluza}  T. Appelquist, A. Chodos, P.G.O. Freud. ''Modern
Kaluza-Klein Theories'', Addison-Wesley, 1987.

\bibitem[6]{ti}  A. Sandoval-Villalbazo, L.S. Garc\'{i}a-Col\'{i}n.
Phys. of Plasmas 7 (2000) 4823-4830.

\bibitem[7]{Stephani}  H. Stephani ''General Relativity'', 2nd. Ed.
Cambridge Univ. Press, 1990.
\end{thebibliography}
\end{document}